# An Analysis of Structured Data on the Web

Nilesh Dalvi       Ashwin Machanavajjhala       Bo Pang
Yahoo! Research
4301 Great America Parkway
Santa Clara, CA 95054
{ndalvi, mvnak, bopang}@yahoo-inc.com

## ABSTRACT

In this paper, we analyze the nature and distribution of structured data on the Web. Web-scale information extraction, or the problem of creating structured tables using extraction from the entire web, is gathering lots of research interest. We perform a study to understand and quantify the value of Web-scale extraction, and how structured information is distributed amongst top aggregator websites and tail sites for various interesting domains. We believe this is the first study of its kind, and gives us new insights for information extraction over the Web.

## General Terms

Experimentation, Measurements

## Keywords

Structured Data on the Web, Information Spread, Information Connectivity

## 1. INTRODUCTION

One of the grand research challenges in the field of information extraction (IE) is to develop effective techniques for Web-scale information extraction. Traditional IE techniques considered in the database community tend to be *source-centric*, i.e., they can only be deployed to extract from a specific website or data source. However, a range of *domain-independent* techniques have emerged recently [2, 4, 9, 10, 11, 16, 20] that seek to look at extraction holistically on the entire Web.

Our own motivation for this work comes from our research goal of building a Web of concepts [7], where we want to extract, link, and organize entities across a large number of domains. There are some domain-independent efforts, e.g. WebTables [4, 10], that extract all simple tables and lists from the Web and store them as relational data. However, domain-independence makes it difficult to attach semantics to the extracted data. Furthermore, the amount of information that can be extracted is quite limited, since most of the information is outside of simple tables and lists. In contrast, we have advocated a *domain-centric* approach to the problem, where we want to extract all the entities and their attributes from the entire Web restricting to a specific domain [7]. For instance, one might be interested in constructing a database of all restaurants, along with their contact information, hours of operation and set of reviews, by extracting information from all the websites on the Web that belong to the restaurant domain. Similarly, one might be interested in constructing databases of all books and their reviews, artists and their discography, or product listings from merchants, and so on.

There are several aspects to a complete solution for this end-to-end challenge, which include automatic crawling, clustering, extraction, deduplication and linking, all at the scale and diversity of the Web. Some of these problems have been well-studied, e.g. crawling shallow-web [5] and deep-web [18]. Others, like unsupervised site extraction [1, 6, 8, 15, 16, 17, 20], have received lots of attention recently, and are a topic of active research.

In this paper, we consider a fundamental problem underlying all these techniques, which is discovering all the sources of structured information on the Web that contain entities in the domain of interest. Understanding the distribution of structured data is crucially important in order to evaluate the feasibility and efficacy of web-scale extraction. Although the problem of web-scale extraction is gathering interest, does the problem exist in reality? What if there exists a definitive source (or a small number of sources) that can be manually wrapped to construct complete databases for the domain of interest? Or, what if the unpopular tail entities that are not mentioned in definitive sources are so invaluable for users that it is not worth extra efforts to discover or extract structured information about them? By using Yahoo! data available to us that includes a crawl of the Web, a database of structured business listings from Yahoo! Local, and Web search logs, we perform a first of a kind large-scale analysis that attempts to quantify the distribution of domain content on the Web for various domains (like books, restaurants, etc.) as well as the *value* of extracting structured information from the *whole* Web.

We believe our study sheds several new insights to the problem of web-scale information extraction. We summarize our main conclusions here:

1. First, we observe that even for domains with well-established aggregator sites (e.g., `yelp.com` for restaurants) we need to go to the long tail of websites to build a reasonably complete database of entities in the domain. This provides additional motivation for the problem of *domain-centric information extraction*. Many of the current efforts on web-scale extraction are domain-independent, in that they try to extract "*everything from everywhere*", across all domains. In contrast, given our observations, we believe that the problem is challenging even when restricted to a specific domain on the Web. We quantify some of these challenges in this study.

2. Second, there is significant value in extraction for the long tail. While both the demand for entities and the availability of information about the entities decay as we move towards the tail, information availability decays at a much faster rate. This suggests there can be higher value in tail extraction in spite of the lower demand in tail entities.

3. Third, we observe that the aggregate content within a domain is well-connected, and there is a significant amount of content redundancy. While this finding is not very surprising, to the best of our knowledge, this is the first study that quantifies the connectivity and redundancy of structured data on the Web. We believe that structural redundancy within websites, content redundancy across websites, and entity-source connectivity together can be leveraged to develop effective techniques for *domain-centric information extraction*. Given that the general domain-independent extraction with really high precision and recall seems out of reach of the current state of the research, solving domain-centric extraction might be an important first-step in this direction.

## 2. SCOPE OF ANALYSIS

We analyze data sources on the Web to address two main questions. First, we examine what is the value of extracting from the *whole* Web as opposed to focusing on popular sources and popular entities. We then study connectivity properties of these data sources with implications on whether it is possible to discover all data sources easily and quickly. Our experiments are organized into three topics:

- **Spread of data among sites.**

    We start by examining whether winners *cover* it all. For each domain, there are usually a small number of top sites, each covering a large and seemingly comprehensive list of entities. Does this mean one only needs to identify and wrap a few top sites in order to build a comprehensive database? What if we want some redundancy in the data sources to overcome errors introduced by a single source (e.g., mistakes in the underlying database or noise in the extraction)? Using several different domains, we study how the information about various attributes is spread between top head sites and a large number of tail sites. Our surprising finding is that, even for domains with strong head aggregators, we need to go to the long tail of websites to build a complete database of entities. For instance, if we want to extract the homepages of restaurants, we need to access at least 1000 websites to get a coverage of 90%, and we need to access at least 5000 websites to get 95% of all restaurant reviews on the Web. In addition, requiring redundancy can drastically increase the number of sites needed to reach reasonable coverage. Section 3 describes the methodology and results in details.

- **Value of tail extraction.**

    We then study whether there is a substantial need to construct a comprehensive database: for instance, if restaurant or a book currently has no reviews listed in a head site, maybe this is because no one cares about it in the first place? We take 3 popular sites (each from a different domain) and study the relationship between demand and availability of review content, where we measure the demand for an entity by the number of visits to the entity page on the site. While it is true that entities with less available content are also less visited, we find that the decay in content availability is faster than the decay in demand in at least two domains. This means that extraction for unpopular tail entities can have bigger value for the user base, even though they are relevant to a smaller group of people. Section 4 describes the methodology and results in details.

- **Diameter and Connectivity of data.**

    Once we establish the need to extract from a large range of sources, the next natural question to ask is whether these data sources can be easily discovered. The answer to this question is dependent on the actual algorithm used. We consider a general class of bootstrapping-based algorithms, where one starts with seed entities, use them to reach all sites covering these entities (for instance, via search engines), expand the set of entities with all other entities covered on these new sites, and iterate. Rather than conducting experiments with specific algorithms, we instead study the property of the entity-site bi-partite graph, which shed light on the upper-bound performance for any algorithm in this class. Clearly each seed entity can only reach sites in its connected component. Connectivity properties have implications in how robust such algorithms will be with respect to the choice of the seed set, and diameters indicates how many iterations are necessary. We find that the entity-site graphs are highly connected across many different domains – the largest connected components cover well over 99% of the graph, and diameters are reasonably small. Furthermore, these graphs are still highly connected even after top sites are removed, indicating an additional layer of robustness. The details can be found in Section 5.

## 3. SPREAD OF DATA ON THE WEB

In this section we analyze how structured information on the Web is distributed for a broad range of domains. Consider an example domain of *Restaurants*. Suppose we want to extract attributes like *name*, *address*, *phone*, *homepage* and *reviews*. On one end of the spectrum, there are large-scale global websites like `yelp.com` and `urbanspoon.com` that contain structured data about a large number of restaurants. On the other end, there are local aggregator pages, city chambers of commerce websites, or even individual critics blogs, each covering a small number of restaurants of local interests. We want to answer the following question: how is the information distributed between the top head sites and the large number of tail sites? What is the recall if we identify and extract from the top-k websites in the domain?

At first glance, the problem seems as hard as the actual problem of extracting information from the entire web. To answer the question exactly, we need to identify all the websites containing information of interest, including the long tail of small aggregator sites. Furthermore, we need full-blown extraction components for all possible attributes across different sites. Since the entire purpose of the analysis is to motivate the problem of Web-scale extraction, this approach is infeasible. Thus, we address an alternative question. Instead of attempting to study the distribution of structured information for *all* attributes of *all* entities, we examine the distribution for *some* attributes of *a reasonably comprehensive set of* entities.

### 3.1 Methodology

We avail of the various datasets that we have at Yahoo!. We look at domains with the following two properties:

1. We already have access to a large comprehensive database of entities in the domain.

2. The entities have some attribute that can uniquely (or nearly uniquely) identify the entity, e.g., phone numbers of businesses and ISBN numbers of books.

At Yahoo!, we have such datasets for several domains, e.g., *restaurants*, *shopping/retail stores*, *schools* and *libraries*. We make use of our *Web cache* data, which contains all webpages crawled by Yahoo! search engine. For each domain, we go through the entire Web cache and look for the identifying attributes of the entities on each page. We group pages by hosts, and for each host, we aggregate the set of entities found on all the pages in that host.

Note that we have reduced the problem of analyzing the spread of data on the web to a task that is much easier than actual web-scale extraction. First, we are not extracting new entities here, but only studying the distribution of information about entities we already have in our database. Second, we use the identifying attributes of the entities to establish the presence of entities on webpages. Nonetheless, we believe our findings can still shed some light on the original question. See Section 3.5 for more discussions on the implications of the methodology we adopted.

### 3.2 Data

We look at 9 different domains. A list of domains, along with the attributes studied is given in Table 1.

For the `books` domain, we used a database of `ISBN` numbers of all books published before 2007. The database has around 1.4M entities. To identify `ISBN` numbers in webpages, we look for matches to one of the `ISBN` numbers from our database, formatted either as a 10-digit or a 13-digit `ISBN`, along with the string "*ISBN*" in a small window near the match. For the other 8 local business domains listed in Table 1, we used the Yahoo! Business Listings database, which has a comprehensive list of entities in each domain.

| Domains | Attributes |
|---|---|
| `Books` | `ISBN` |
| `Restaurants` | `phone, homepage, reviews` |
| `Automotive` | `phone, homepage` |
| `Banks` | `phone, homepage` |
| `Libraries` | `phone, homepage` |
| `Schools` | `phone, homepage` |
| `Hotels & Lodging` | `phone, homepage` |
| `Retail & Shopping` | `phone, homepage` |
| `Home & Garden` | `phone, homepage` |

Table 1: List of Domains

The database contains millions of business listings in the US. While the exact size of this database is suppressed as the information is proprietary, note that it is sufficiently large for the purpose of this study. We looked at two attributes: *phone numbers* and *homepage URLs*. For phone numbers, we used a standard regular expression based US phone number extractor. For homepages, we looked at the content of `href` tags of all anchor nodes in pages. Finally, we considered an additional attribute of *reviews* for the restaurant domain. To detect whether a given webpage contains a review of a given restaurant, we took all pages on the Web containing a matching restaurant phone number, and used a Naïve-Bayes classifier over the textual content to determine if a page has review content.

### 3.3 Metrics

We analyze the distribution of data using the following metrics. For a given domain, let $N$ be the number of entities in the database. For each website on the Web, we find the set of entities present in that website. Then, we order the list of websites in decreasing order of the number of entities they contain.

Given a set of websites $W$ and a positive integer $k$, we define the $k$-coverage of $W$ as the fraction of entities in the database that are present in at least $k$ different websites in $W$. Thus, 1-coverage is simply the number of unique entities in $W$. Analyzing $k$-coverage for $k > 1$ is interesting because one may be looking for a piece of information from $k$ different sources to place a high confidence in the extraction.

For each $t$, we look at the top $t$ websites as ordered above, and plot $k$-coverage ($1 \le k \le 10$) of these $t$ sites as a function of $t$. These plots reflect how spread out is the distribution of information for a given domain.

### 3.4 Results

Figure 1 and Figure 2 contain the resulting coverage plot for the phone attribute and the homepage attribute for the 8 local business categories. Figure 3 contains the coverage for book ISBN numbers. Figure 4 shows the coverage plot for restaurant reviews.

We start by analyzing the plots in Figures 1–3. The 10 different curves in each plot, from top-left to bottom-right, correspond to the $k$-coverage for $k = 1$ to 10. Let us look at Figure 1(a) that contains the coverage of phone attribute of restaurants. If we look at $k = 1$, we see that the top-10 sites cover around 93% of all the entities in the database, and if we extract from top-100 sites, we get close to 100% coverage. If we want at least $k = 5$ pages for each phone

682

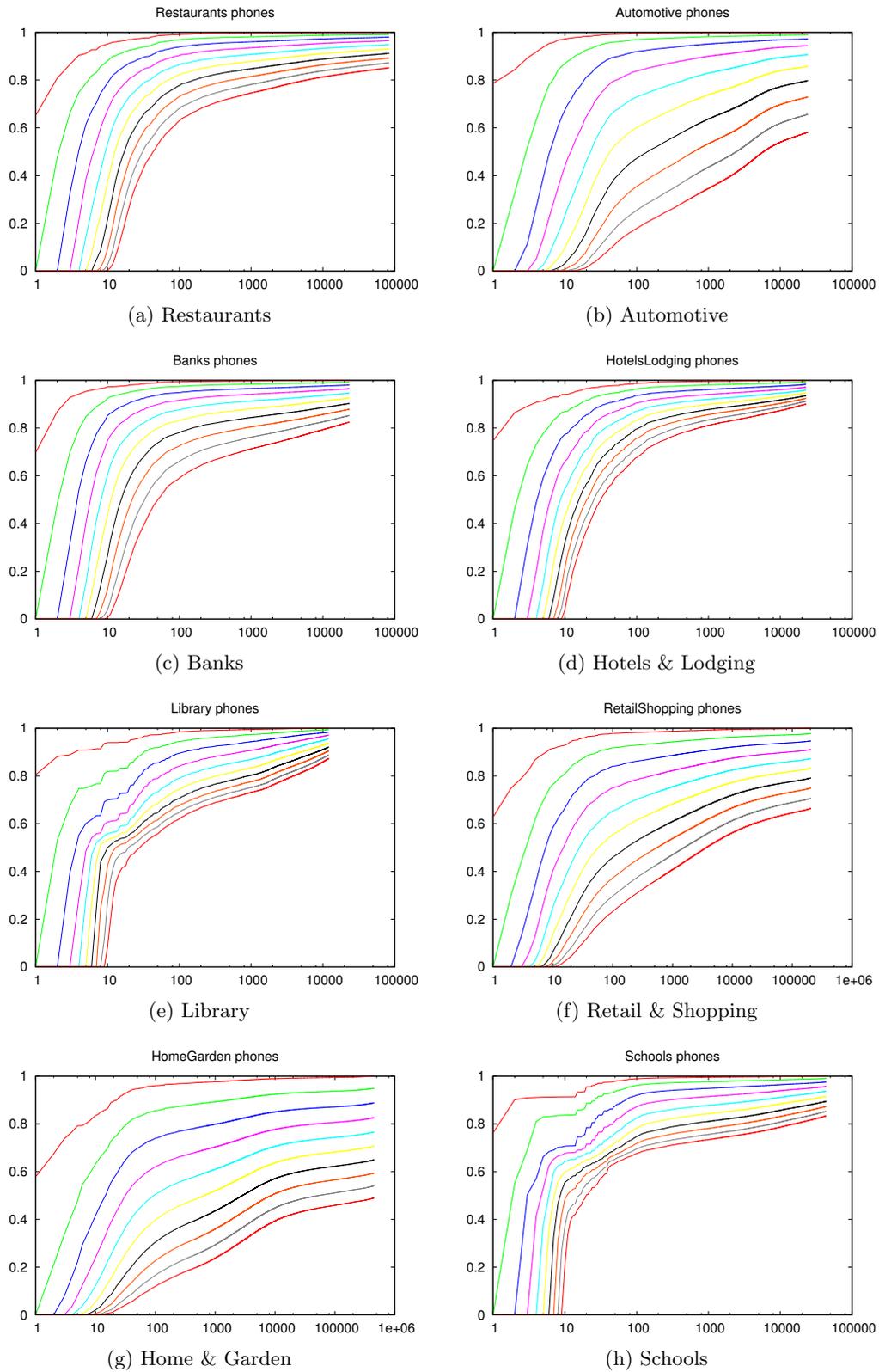

Figure 1: Spread of Phone Attribute for Various Domains. The 10 curves in each graph, from top-left to bottom-right, correspond to $k$-coverage for $k = 1$ to $10$.

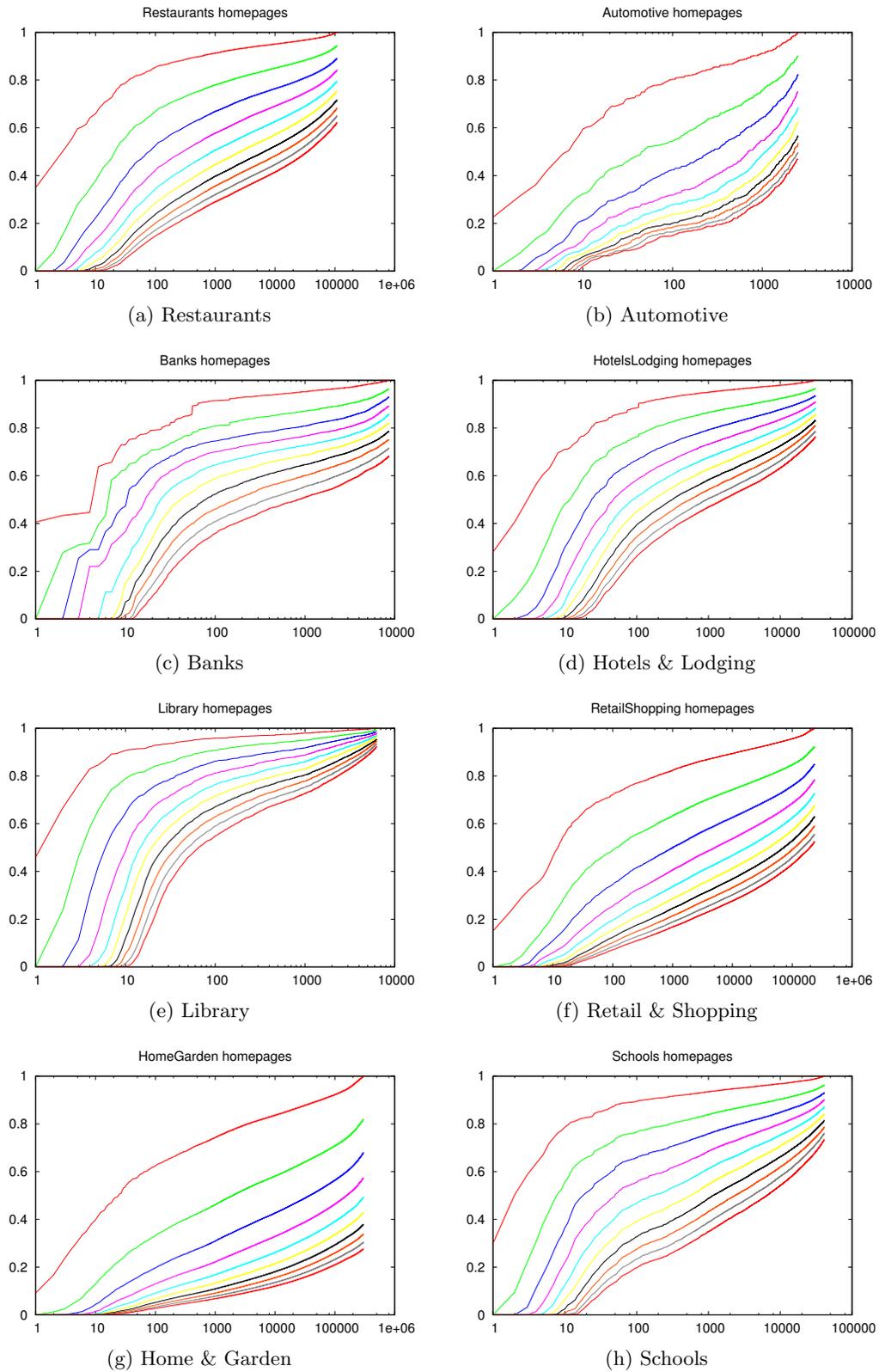

Figure 2: Spread of Homepage Attribute for Various Domains. The 10 curves in each graph, from top-left to bottom-right, correspond to $k$-coverage for $k = 1$ to $10$.

684

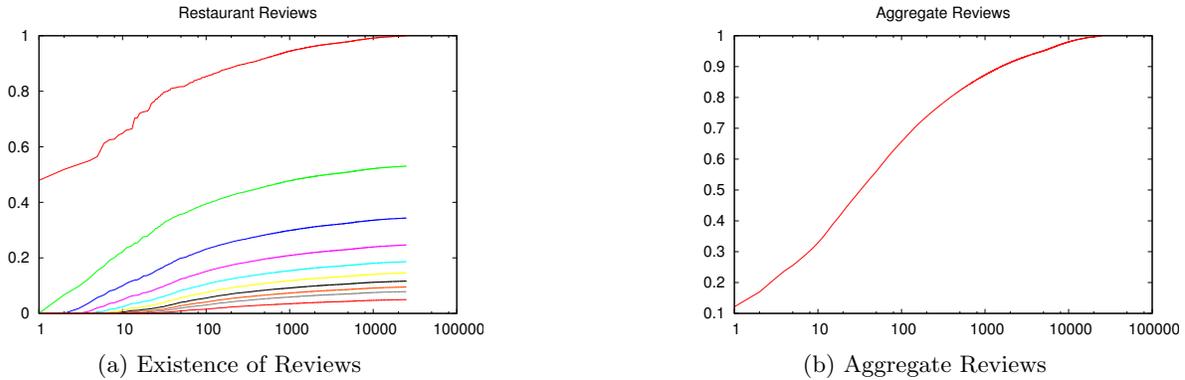

(a) Existence of Reviews
(b) Aggregate Reviews

Figure 4: Spread of Review Attribute for Restaurants

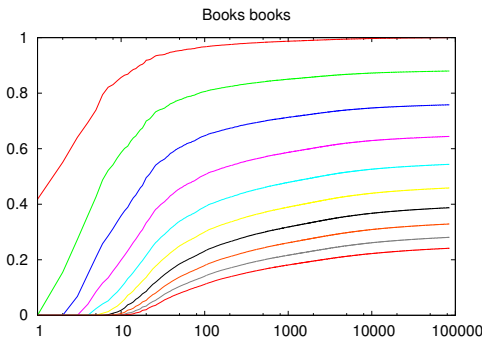

Figure 3: Spread of Book ISBN Numbers

number, then we need to got to top-5000 sites to cover even 90% of all phones.

The graph suggests that the phone attribute in restaurants domain has a high *availability*, and we can obtain a nearly complete database by manually extracting from top-100 sites (which is still a lot of manual effort). Though if we want our extractions to be corroborated from at least k pages, we need to extract from a much larger set of websites to achieve high coverage (more that 5000 sites for k=5). Similar trends can be observed in other local business domains as well as for the ISBN attribute of the book domain. In fact, the gap between curves corresponding to different $k$ values can be even bigger.

As we move towards less available attributes from the same domain, the spread starts to increase even for $k = 1$. Consider Figure 2(a), which plots the coverage of the homepage attribute in the restaurant domain. We see that spread here is much larger than what we observe in Figure 1(a). We need at least 10,000 sites to cover 95% of unique restaurants (even with $k = 1$). The plots for phones and homepages for other domains tell a similar story.

We next look at the availability of restaurant reviews on the Web. Review attribute differs from phones and homepages in an interesting way, as there can be multiple reviews of the same restaurant, and each review adds value to the restaurant. Thus, we can define the review coverage in two ways. First, we say that a website $w$ *covers* a restaurant $r$ with respect to reviews if $w$ has at least one (webpage containing a) review about $r$. We define the $k$-coverage of a group of websites $W$ as the fraction of restaurants covered by at least $k$ sites in $W$. Figure 4(a) plots the $k$-coverage of reviews. Again the curves from top to bottom correspond to $k$-coverage ($k = 1, 2, \ldots, 10$). If we want at least 1 review, we need to extract from more than 1000 sites to get 90% coverage, and from more than 5000 sites if we want to reach 90% coverage of entities with reviews from at least 2 sites.

A second way to define review coverage is to look at the total number of all the webpages on the Web that contain a restaurant review. Then, we can look at the fraction of those webpages covered by the top-$n$ sites as a function of $n$ (unlike $k$-coverage, which can be defined for each $k$, we only have a single curve here). Figure 4(b) plots the resulting curve. We see that the spread here is higher. E.g. the top 1000 sites cover around 95% of unique restaurants in Figure 4(a), but only cover 80% of the total reviews on the Web in Figure 4(b).

In conclusion, we observe that tail sites carry a significant amount of information, even for domains like *restaurants* that have well-established head aggregator sites.

### 3.4.1 Ordering Sites by Diversity

In the coverage analysis, we ordered the sites by their individual coverage. However, the top-k sites may have a huge amount of duplicated content. So, it may be the case that while very few sites are sufficient to cover all the entities, they may have to be chosen carefully. To this end, we consider the following experiment[1]: for a given $k$, we try to find the set of $k$ sites that together give the maximum coverage, and study this as a function of $k$. Stated as such, this is the classic set cover problem which is known to be NP-hard. To make this experiment computationally feasible, we consider the best known approximation algorithm for set cover, which is the *greedy approximation algorithm*. In this algorithm, sets are chosen greedily such that at each step, the set that contains the largest number of uncovered elements is chosen.

Figure 5 shows the resulting coverage plot for the homepage attribute of restaurants. It compares the 1-coverage of top-$t$ sites as selected by the greedy set cover algorithm with the 1-coverage of top-$t$ sites ordered by size (i.e., the $k = 1$ curve in Figure 2(a)). While the coverage slightly improves with the greedy set cover, the improvement is insignificant. The plots for other attributes and domains are omitted as they tell the same story. We conclude that a careful choice of hosts does not lead to significant increase in coverage by top sites.

---

[1] We thank the anonymous reviewers of this paper for suggesting this experiment.

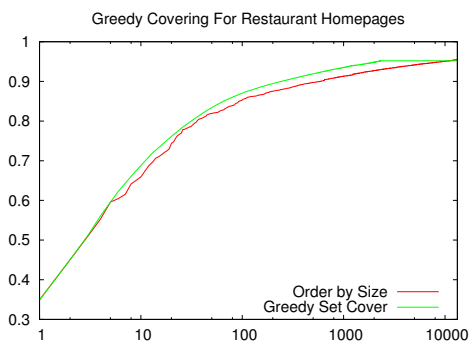

Figure 5: Ordering Sites by Diversity

## 3.5 Discussion on Errors in Methodology

Since any experiment of this nature (short of fully solving the grand challenge of web-scale information extraction) is prone to errors, we identify potential sources of errors here and discuss their implications. We see two sources of errors that may arise in our methodology. The first is that we are only analyzing how the entities from our dataset are spread over the Web, and using it as an approximation for the true distribution of all the entities from the corresponding domain on the Web. However, since we have sufficiently large datasets, we believe the approximation error is insignificant. That is, we believe that the scale of the datasets is large enough for us to assume that the data distribution, restricted to our datasets, reflects the true distribution. (And when it differs, it would most likely over-estimate the coverage by head sites, as it under-estimate how much is missing from them.) A second potential source of error is the false matching of identifying attributes. For instance, a page might contain a 10-digit number, which is not meant to be a phone number, but happens to be formatted in the right way and matches a phone number in our database. Based on small random samples, we observed that the regular expression matching of US phone numbers, URLs and ISBN numbers had a high accuracy. Furthermore, for websites with multiple matches, it is even less likely that the matches are accidental. Even if false matches do creep in, they will only lead to over-estimation of the coverage (i.e., making the spread appear lower), since the top-$t$ websites will report more entities than what they truly cover. Thus, it only strengthens the conclusion that a significant amount of information can only be found in the tail.

## 4. VALUE OF TAIL EXTRACTION

We have shown that in order to build a complete database for a given domain, we need to extract from a large number of sites. However, one would expect (especially in a Web-setting) that a significant fraction of queries issued to the database would pertain to a small fraction of popular (head) entities than for unpopular (tail) entities. These head entities may be extracted from a small number of top aggregators. Thus one might conclude the additional *value* gained by including tail entities is little compared to the extra cost of extracting from an order of magnitude more sites, diminishing the importance of building a complete database and that of Web-scale extraction. In this section, we argue to the contrary and experimentally analyze the value of extracting information for tail entities.

First, we observe that queries on databases not only request entities but also their attributes; and as shown in the previous section, sites vary in their coverage of attributes. Attributes of entities can be classified into two categories – *closed* and *open*. Closed attributes are those that we expect to have one single correct value (most of the time), e.g., phone number of a local business and manufacturer of a product. If we ignore the need for redundancy to ensure correctness, once one value is found for that attribute, there is no need to look further. On the other hand, *open attributes* are set valued whose sizes are not known a priori. Consider, for instance, a list of available vendors for a product or user generated reviews for an entity.

While the argument for the importance of extracting closed attributes for tail entities is similar to that of finding tail entities themselves, the value in extracting open attributes is more interesting. Compared to the first few values for an open attribute, the $k^{th}$ value provides less additional information for that attribute. For instance, consider an open attribute like user generated reviews. We expect head entities to already have many reviews from top aggregators. Hence an additional review might provide little additional information, though it is of interest to many users. On the other hand, we expect a tail entity to have few or no reviews on top aggregators. While an additional review is very informative in this case, it might not have much user demand. Hence, an interesting question arises for open attributes: what is the *value* of adding a new value of an open attribute for head vs tail entities?

In the remainder of this section, we address the above question via an experimental case study of the value of extracting user generated reviews for tail entities in three domains. Section 4.1 describes two different ways of measuring user demand empirically, and introduces the datasets constructed for three representative domains with potentially different demand patterns. Section 4.2 examines the demand patterns in more detail, and Section 4.3 addresses the main question: what is the value-add of a new user review for head vs. tail entities?

## 4.1 Data

We approximate user demand from two ways of estimating Web traffic: one year of user search traffic on Yahoo! Search (*search*) and one year of user browsing activities recorded by Yahoo! Toolbar (*browse*). In both cases, we extracted user clicks on URLs that correspond to a unique structured entity. We selected the following three sites that have high user traffic and are also rich in user generated reviews.

- Amazon: This site is rich in products. We focused on URLs matching the pattern amazon.com/gp/product/[ID] or amazon.com/*/dp/[ID] and used the 10-digit product ID encoded in the URL as the key for the entity. We picked a random sample of over a million such pages.

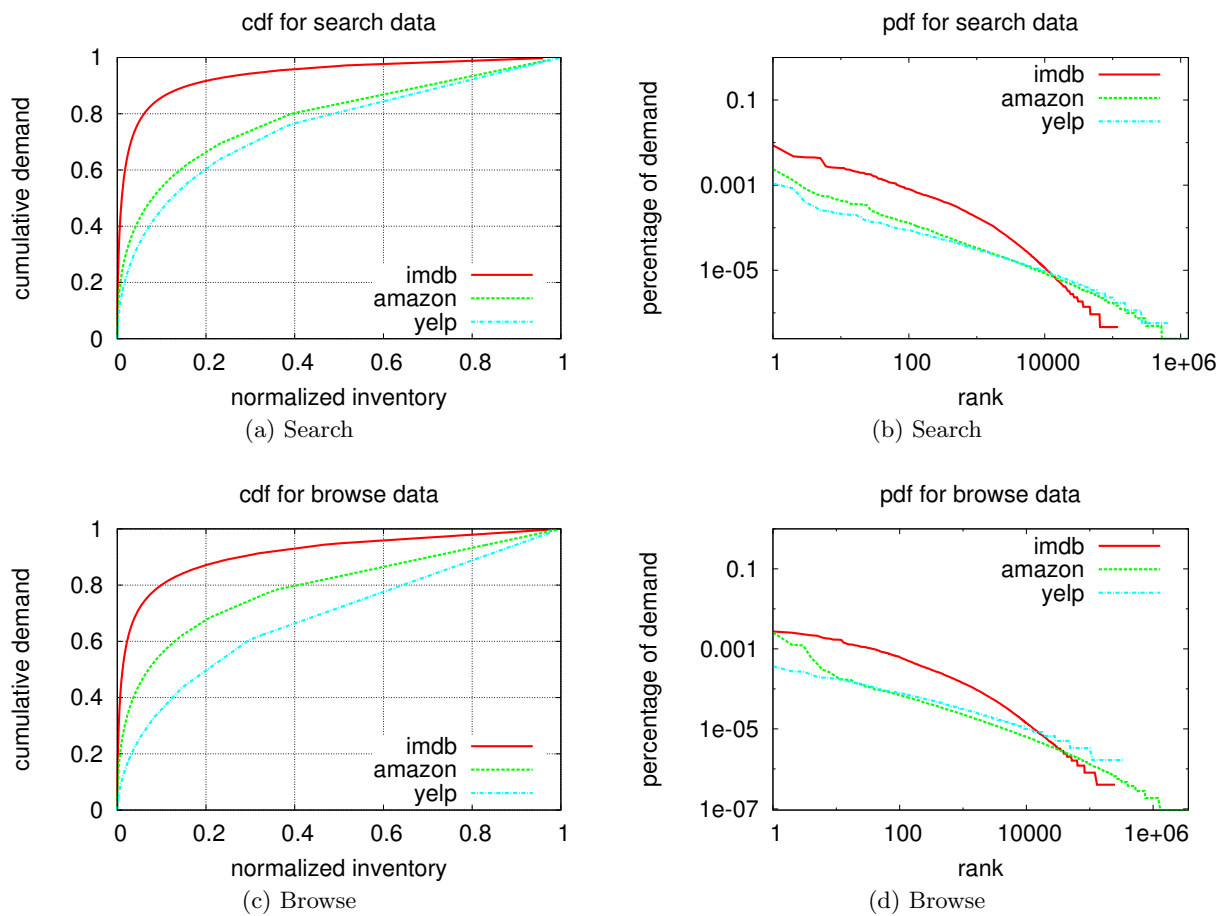

Figure 6: The long tail of demand in different domains: aggregate view

- Yelp: This site is rich in local business entities. We extracted entity IDs from URLs matching `yelp.com/biz/[ID]`. We used a sample of over 500K entity pages from Yelp.

- IMDb: We extracted movie entity IDs from over 100K URLs matching `imdb.com/title/tt[ID]`.

We use unique (anonymized) cookies as a proxy for unique users, and define the demand for a URL (and hence the entity it mentions) as the number of visits from unique cookies[2]. The browse data simulate the typical traffic patterns observed on a site (subject to the sampling bias of limiting to toolbar users). Traffic may be either biased to tail items as a result of personalized recommendation on the site, or biased towards head items as a result of popular items presented earlier in intra-site search results. On the other hand, traffic in the search data simulate the raw demand since these were the items users actively searched for.

### 4.2 Demand

Figure 6 summarizes the overall demand patterns for the three domains, and the overall trends are similar in both the search and browse data. Figure 6(a) and 6(c) present cumulative demand satisfied as a function of the percentage of overall inventory. In Figure 6(a), while top 20% of movie titles account for more than 90% of the overall demand on IMDb, top 20% of business entities account for only 60% of the overall demand on Yelp. The difference is even more pronounced when demand is measured in the browse data (Figure 6(c)).

We make two observations. First, the demand patterns in the three domains are different. The demand curve for Yelp is the flattest while that for IMDb is the sharpest. This is expected: a top movie title can be watched by millions of people at the same time, whereas even the most famous restaurant can only serve a small number of clients in one day; demands for products sold on Amazon is more spread-out than movies but is still more concentrated than local business entities.

Second, at least for products and local business entities, the demand curve is heavy-tailed. While for movies head entities account for most of the demand, note that there is a subtle difference between satisfying a significant portion of the demand (e.g., accounting for most of the observed consumption) and satisfying a significant portion of the users. A recent study examined the value of tail entities from the user perspective [13] and found that nearly every user had some niche interests represented in the tail, even though these tail entities may only account for a small fraction of the total demand. For instance, while tail entities in Netflix and Yahoo! Music[3] account for only 13% and 34% of total ratings, 90% of Netflix users and 95% of Yahoo! Music users have rated tail entities at least once, and 35% of Netflix users and 70% of Yahoo! Music users regularly do so [13]. That is, satisfying 90% of the users 90% of the time requires a better coverage over tail entities than what is necessary to satisfy 90% of total consumption.

---

[2]In the search data, we took the number of unique cookies per month; in the browse data, we took the number of unique cookies in the year.

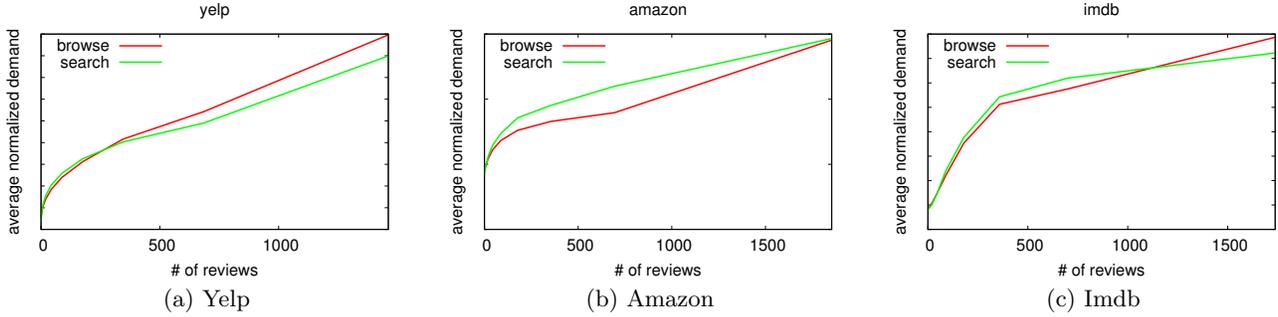

Figure 7: Normalized demand vs. number of existing reviews. Demand (number of visits) measured in the browse data and the search data are normalized within each dataset to have a mean of zero and standard deviation of one.

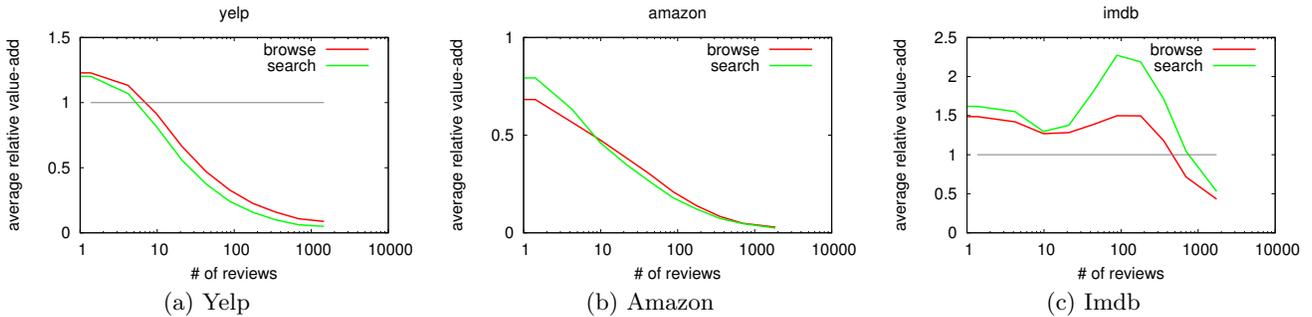

Figure 8: Average relative value-add ($\frac{VA_{(n)}}{VA_{(0)}}$) of adding one review as a function of the number of existing reviews ($n$). The relative value-add for objects with zero reviews is one, and is shown in the plots for reference.

While the results in this section are not surprising, when we compared the demand curves with the availability curves, we did get surprising results, as we describe next.

## 4.3 Demand vs. Available Content

As mentioned earlier, popular entities have many reviews in top aggregator sites. As we move towards the tail, both the amount of available content and the demand for such content taper off. Hence, while extracting an additional review for a tail entity increase the information value per user, there are fewer users interested in such entities. We quantify the value of adding a new review *VA* as the product of the demand and the *average additional information* $I_\Delta$ provided by that review. We first describe our choice of $I_\Delta$, and then present our experimental study on *VA* for different domains.

---

[3] *Tail* is defined as items not available in large, bricks-and-mortar retailers in [13], which is slightly different from the way we are using this term, where it is defined based on percentage of the overall inventory. The tail items from [13] also satisfy our definition of tail.

### 4.3.1 Additional Information of a New Review

Consider an entity that has $n$ reviews. We would like to construct a function $I_\Delta(n)$ that quantifies the average additional information provided by the $(n+1)^{th}$ review. While it is hard to know the exact form of this function, intuitively, we expect it to be inversely related to $n$. In our experiments, we use $I_\Delta(n) = 1/(1+n)$. An inverse linear function is motivated by aggregation scenarios. Suppose reviews are shown to the user in a summarized form (e.g., an average sentiment polarity for each aspect of a product), and the summary considers each review as an independent source of true information. For instance, if an entity has $n$ reviews all giving a "thumbs-up" for ambiance, if the next review gives a "thumbs-down", this would result in the biggest change in the overall rating, but even in this case, it would impact the overall rating only by an additive factor of $1/(1+n)$. Thus, $I_\Delta(n)$ bounds the influence the $(n+1)^{th}$ review can have on the average presentation.

Let $k$ be the demand for this entity $e$. We define the *value-add* of a new review to $e$ as $k/(1+n)$, the product of the demand for the entity and $I_\Delta(n)$. That is, if the additional information provided by the review is quantified as $I_\Delta(n)$, and this information is of interest to $k$ users, the total additional value for all users in $k \cdot I_\Delta(n)$. We compute the average value-add for all entities with $n$ reviews, and denote it as $VA(n)$.

One could also consider alternatives for $I_\Delta$ based on other typical modes of review consumption. For instance, $I_\Delta(n)$

could be a step function that gives zero weight when $n \geq c$ for a small constant $c$ (like 10). This captures the scenario where a user reads no more than $c$ reviews. Such alternatives correspond to a more dramatic decay of $I_\Delta$ for head items, and only serve to strengthen our main message of this section: compared to our choice of $I_\Delta$, these alternative choices would estimate even higher value-add of extracting a new review for tail entities.

In reality, the value also depends on the quality of reviews, but to facilitate such analysis, we will necessarily have to assume the quality of review is independent of the order of discovery.

### 4.3.2 Value-Add For Head vs. Tail Entities

Figure 7 plots the average browse and search demand across entities that have $n$ reviews as a function of $n$. Thus, the tail entities lie in the beginning of the curve (small $n$) and head entities lie at the end (large $n$). There indeed is more demand for entities with more reviews.

Figure 8 plots the normalized value-add, $VA(n)/VA(0)$, as a function of $n$, using either search or browse data to estimate demand. A value of $VA(n)/VA(0) \geq 1$ implies that the average demand for entities with $n$ reviews is at least $(n + 1)$ times larger than the average demand for entities with no reviews.

Since for larger values of $n$ there are fewer entities in each $n$-group, we grouped entities based on the value of $log(n)$.[4] We see that for both Yelp and Amazon, value-add decreases as $n$ increases. That is, one additional review for a head item has lower value-add than an additional review for a tail entity (note that the left end of the curve corresponds to the tail and the right end corresponds to the head). Since value-add is the ratio between the demand $k$ and availability $n$, this means demand does not increase as fast as content availability. In other words, for both of these domains, as entities gets less popular, the available content decays faster than user interest. For the IMDb data, the relative value-add goes up for entities with mid-range popularity but then falls off for the head entities — this could be due to a more drastic decay in user interest for tail entities.

Results shown here may appear counter-intuitive: one might assume that demand of a product ($k$) is proportional to the number of users who buy it, which, in turn, is proportional to the number of people who write reviews about the product. In such a case, we should observe curves close to $y = 1$ in Figure 8. There are two possible explanations: first, what we measure here are researching demand (i.e., users who are browsing or searching for an entity) rather than transactional demand, and the relationship between the two is non-trivial to model. Indeed, it could be that a higher percentage of users who are viewing / searching for a popular item end up purchasing the product or patronizing the restaurant (i.e., a higher conversion rate for popular items). This can lead to the trend that we observe in Figure 8. Note that $k$, rather than actual transactional demand, is the demand of interest to us, since our goal is to assist any user who wants to consider the entity.

Second, the relationship between the number of people who purchase a product and the number of people who write a review is complex. While less popular entities might be in a greater need of a fresh review, potential reviewers could be encouraged by existing reviews for popular entities and choose to contribute to an existing micro-community. There are separate studies looking into what motivates reviewers to write the $1001^{th}$ review for a given product [12]. An in-depth treatment of this subject is beyond the scope of this paper.

---

[4]That is, entities with 0 reviews form the first group, entities with 1-2 reviews form the second, and so on. Entities with 1023 or more reviews form the final group.

In summary, as we move towards the tail, the availability of content decays much faster than the demand for content. In other words, even when adjusted for a smaller demand, the lack of information for tail entities is still more prominent, and in that sense, extraction for tail entities can bring a higher value add to the user base.

## 5. CONNECTIVITY OF STRUCTURED INFORMATION

An important property of the structured data on the web that we want to analyze is whether the data sources are well "connected" by overlapping entities. In order to discover and extract from all possible sources on the Web, most techniques rely on content-redundancy, i.e, they require content to overlap between the sources which the system has already discovered, and the remaining (yet to be discovered) sources on the Web. These include techniques like Flint [3], Knowitall [11] and several set expansion techniques [14, 19, 21], which iteratively increase the coverage of their extraction by using information extracted in one iteration to discover new sites and then extracting from these in the next iteration. We model this iterative discovery of new entities and new websites by constructing a bi-partite graph, whose nodes are entities and websites, and there is an edge between an entity and a website if the website covers the entity.

For each of these graphs, we then analyze:

- The number of connected components of the graph.
- The size of the largest component.
- The diameter of the graph.

### 5.1 Entity-Website Graph

We consider a bipartite graph between the set of entities in a given domain and the set of websites, where there is an edge between an entity $e$ and a website $h$ if there is a webpage in $h$ that contains $e$. To determine if a given webpage contains a given entity, we use the same methodology that we employ in Section 3 and look for identifying attributes of entities on webpages. For each domain and attribute listed in Figure 1, we draw a corresponding graph using the corresponding attribute. Table 2 shows the average number of sites mentioning an entity in each of these graphs.

### 5.2 Diameter

The diameter $d$ of a graph is defined as the length of the longest path among shortest paths between all pairs of nodes in the graph. Thus, every pair of nodes is connected by a path of length at most $d$. From an extraction perspective, the diameter of the structured entity-website graph is important for the following reason. Suppose we start with a small set of seed entities. At each iteration, we discover all the sites that contain entities overlapping with the current set of entities, and then extract all the entities from these sites, and add them to the current set. Given such a "perfect" set expansion algorithm, starting from any seed set,

| Graph | | Avg. #sites | diameter | # conn. | % entities in |
| Domain | Attr | per entity | | comp. | largest comp. |
| --- | --- | --- | --- | --- | --- |
| Books | ISBN | 8 | 8 | 439 | 99.96 |
| Automotive | phone | 13 | 6 | 9 | 99.99 |
| Banks | phone | 22 | 6 | 15 | 99.99 |
| Home | phone | 13 | 8 | 4507 | 99.76 |
| Hotels | phone | 56 | 6 | 11 | 99.99 |
| Libraries | phone | 47 | 6 | 3 | 99.99 |
| Restaurants | phone | 32 | 6 | 52 | 99.99 |
| Retail | phone | 19 | 7 | 628 | 99.93 |
| Schools | phone | 37 | 6 | 48 | 99.97 |
| Automotive | homepage | 115 | 6 | 10 | 98.52 |
| Banks | homepage | 68 | 8 | 30 | 99.57 |
| Home | homepage | 20 | 8 | 5496 | 97.87 |
| Hotels | homepage | 56 | 8 | 24 | 99.90 |
| Libraries | homepage | 251 | 6 | 4 | 99.86 |
| Restaurants | homepage | 46 | 6 | 146 | 99.82 |
| Retail | homepage | 45 | 7 | 1260 | 99.20 |
| Schools | homepage | 74 | 6 | 122 | 99.57 |

Table 2: Entity-Site Graphs and Metrics

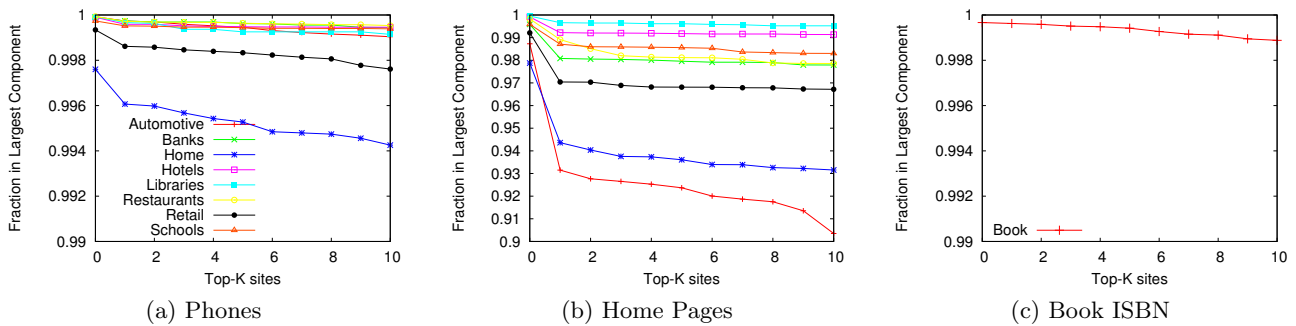

Figure 9: Robustness: fraction of entities in the largest component after top-$k$ sites are removed for local business domains and the book domain.

the number of iterations it takes to extract all the entities is bounded by $d/2$.

Computing the exact diameter of a graph has complexity cubic in the number of nodes in the graph, but can be computed more efficiently when the diameter of the graph is small. We start breadth first traversals from each node in parallel and compute the number of steps needed for each node to discover every other node in its connected component. The values of $d$ thus computed are shown in Table 2. In all graphs we find that $d/2 \leq 4$.

## 5.3 Number of Connected Components

Next, we study the number of connected components in these graphs. Again, iterative bootstrapping-based information extraction can only explore those websites and entities that are in the same connected component as the entities and/or websites in the seed set. Table 2 shows the number of connected components. While some of these graphs, for instance, `Home & Garden - phones` and `Retail - homepages`, have more than a thousand components, in all graphs the largest component covers close to **99%** of all structured entities in the graph. That is, all the disconnected components contain at most one or two entities mentioned only by tail web sites. This means one can extract practically all entities just from the largest component. Moreover, this suggests that any seed set of structured entities will contain, with high probability, at least one entity from the largest component; thus we are all but surely guaranteed to discover and extract most of the entities from random seed sets.

In order to better understand the structure of the graph, we examine the robustness of its connectivity. That is, is the graph being "held together" by one or a few top sites? To answer this question, we re-examine the connectivity of these graphs after removing from them the $k$ largest web sites (sorted by the number of entity mentions). In most domains, while the number of connected components can increase by hundreds just by removing the largest web site, the largest component still contains almost all of the entities. Figure 9 plot the fraction of structured entities in the largest component after removing the top $k$ sites that mention the most entities. We find that even after removing the top $k = 10$ web-sites, the largest connected component still contains $> 99\%$ of all entities for `ISBN` and `phone`, and $> 90\%$ of all entities for `homepages`. That is, even in the absence of the top sites, entities and sources still remain well connected, a highly desirable property for the family of set-expansion-based extraction techniques to achieve good performances.

## 6. CONCLUSIONS

In this paper, we analyzed the nature and distribution of structured data on the Web. Using data available at Yahoo! – databases of structured entities, the Web crawl and search logs – we helped quantitatively answer three important questions regarding domain-centric information extraction. First, we showed that techniques for extraction from the "whole Web" are indeed necessary — for many kinds of information one may have to extract from thousands of sites in order to build a comprehensive database, even when we restrict to a given domain with known popular top sites. Second, we showed that extraction for unpopular tail items can have bigger value for users, even after adjusting for the fact that they are relevant to a smaller group of people. This further motivated discovering and extracting from tail sources of content. Finally, we analyze the connectivity of graphs that connect web sites to the structured entities they mention, in order to understand whether it is possible to discover and extract from all possible sources on the Web. We conclude that the content has high connectivity as well as redundancy, making extraction amenable to set-expansion based techniques.